# Spin Wave – Electromagnetic Nonlinear Interaction to Move RF Limiter Frequency Range to Higher Frequencies

Clifford M. Krowne

Electromagnetics Technology Branch, Electronics Science & Technology Division, Naval Research Laboratory, Washington, DC 20375

**Abstract**

It is shown here that the usual operating frequencies for RF limiters, in the nominal 5, 11 or 15 GHz center frequencies, can be lifted up into the much higher frequencies of the 40 – 60 GHz range by a new formula developed from fundamental spin wave – electromagnetic interactions in magnetic material characterized by a magnetization M.

**Introduction**

Use of magnetic material to provide limiting action to RF signals, a process whereby the signal is chopped off beyond a particular power level, had been deemed very desirable because of the extremely low loss experienced by the RF signal [1] – [16]. This feature arose from the narrow linewidths obtainable in such magnetic materials as yttrium iron garnet (YIG) which is a few Oersteds (Oe), to much higher linewidths of a few hundred Oerseds but still very low in compound ferrite and ferromagnetic materials. It is true that diode limiters are available throughout the entire frequency range, but can suffer from certain timing, switching, and other delays and behaviors, highly undesirable in high performance RF electronic components, making up RF systems. This is particularly noticeable in the higher frequencies.

In order to address the problem of requiring enormous power levels with increasing frequency to obtain limiting action, based on a specific and modest set magnetization value M, here it is shown that using equations which have arisen out of fundamental magnetization theory, a rigorous but simple prescription can be developed. Fundamental magnetization theory encompasses quantum representation of spin S commutation operators, an excellent approximation compared to say exact commutation properties of space and momentum operators, and magnon-magnon, magnon-photon, and other quantum particle interactions. It also involves packaging the single spins into groups such that more macroscopic relations hold and many times easier theoretical treatments are obtained.

Here we are interested in turning our attention toward modifying the stoichiometry of the magnetic material's chemical atomic constituents, to optimize the performance. Possible materials to consider are strontium hexagonal ferrites with a magnetoplumbite structure exhibiting a very high anisotropy field of about 18 kOe and a high remanence to saturation ratio, making it possible to realize mm-wave self-biased circulators [17] – [20] for the 40 – 60 GHz and 35 – 50 GHz bands. They conclude that substituted strontium hexaferrites for the realization of a self-biased circulator around 40 GHz present higher anisotropy field than pure strontium hexaferrites and very high remanent-to-saturation magnetization ratios that making it possible to design a self-biased circulator up to 40 GHz. A recent paper this year on using hexaferrite samarium cobalt composites employing non-conductive binding materials with a view toward isolators and circulators and other linear and nonlinear microwave devices, provides more alternatives [21]. Earlier papers on this subject include work by a number of authors [23] – [28]. Finally, recent papers on frequency selective limiters show the current interest in this area [29], [30].

Because of the large literature covering the fundamental theories of spin behavior and interactions in condensed matter solids [31], whether conductive or non-conductive materials used for RF components, those details can be examined by the reader by going to the general literature. And then there is the literature available that treats the use of such materials for RF components, whose basic features all involve some sort of control of the RF signal, which is a more macroscopic electromagnetic wave propagation in a guiding structure. There are two basic uses or classes of uses for magnetic materials in RF electronic devices: 1) linear behavior which uses the nonreciprocal properties of the permeability of the material gained from its magnetization [32] – [37]; 2) non-linear behavior which relies on fundamental quantum particle interactions, or scatterings. It is the latter category which is addressed in this letter.

### Field Theory of Spin Waves Exchange Effects

When magnetic sample like a ferrite is placed in uniform rf and dc fields, the magnetic moments are all aligned when the sample is saturated and if thermal agitation is neglected. However, including the thermal agitation, adjacent magnetic moments may be pointing in slightly different directions. This will give rise to an increase in the exchange energy,

$$E_{ex} = \frac{1}{2}\mathbf{M} \cdot \mathbf{H}_{ex} \tag{1}$$

For small k-space variation of the magnetization throughout the sample, and $a$ the lattice spacing, compared to the reciprocal lattice spacing,

$$|\mathbf{k}| \ll \frac{2\pi}{a} \tag{2}$$

The exchange field may be set down as

$$\mathbf{H}_{ex} = H_e a^2 \frac{\nabla^2 \mathbf{M}}{M} \tag{3}$$

where $H_e$ is the effective exchange field. For the case of a simple cubic lattice, $H_e$ is given in trems of the usual exchange integral $J$ and the intrinsic atomic or site ionic spin $S$,

$$H_e = \frac{4JS^2}{Ma^3} \tag{4}$$

Associated with the exchange field $\mathbf{H}_{ex}$ is the exchange torque vector $\boldsymbol{\tau}_{ex}$ given by the cross product of the magnetization vector $\mathbf{M}$ and $\mathbf{H}_{ex}$,

$$\boldsymbol{\tau}_{ex} = \mathbf{M} \times \mathbf{H}_{ex} \tag{5}$$

Inserting in (5) the expression for the exchange field, one has

$$\boldsymbol{\tau}_{ex} = H_e a^2 \frac{\mathbf{M} \times \nabla^2 \mathbf{M}}{M} \tag{6}$$

The equation of motion EoM for the magnetization vector $\mathbf{M}$, that is the variation of $\mathbf{M}$ with time is to first order in the damping paratmeter $\alpha$,

$$\frac{d\mathbf{M}}{dt} = \gamma_e \mathbf{M} \times \mathbf{H} + \frac{\alpha}{M} \mathbf{M} \times \frac{d\mathbf{M}}{dt} \tag{7}$$

If one neglects losses for the moment, and just focuses attention on the first torque term, then (7) is streamlined to our magnetodynamics equation (MDE),

$$\frac{d\mathbf{M}}{dt} \approx \gamma_e \mathbf{M} \times \mathbf{H} \tag{8}$$

Or writing the right-hand side term as our torque vector $\boldsymbol{\tau}$ hitting the gyromagnetic ratio $\gamma_e$,

$$\frac{d\mathbf{M}}{dt} \approx \gamma_e \boldsymbol{\tau} \tag{9}$$

Substituting for the general torque $\boldsymbol{\tau}$, with our exchange torque $\boldsymbol{\tau}_{ex}$ from (5), one obtains,

$$\boldsymbol{\tau} \rightarrow \boldsymbol{\tau}_{ex} \tag{10a}$$

$$\frac{d\mathbf{M}}{dt} \approx \gamma_e \boldsymbol{\tau}_{ex} \tag{10b}$$

Our final equation of motion EoM for $\mathbf{M}$ is, retrieving our exchange torque from (6), is

$$\frac{d\mathbf{M}}{dt} \approx \gamma_e \frac{H_e a^2}{M} \mathbf{M} \times \nabla^2 \mathbf{M} \tag{11}$$

### Equation of Motion for Magnetization Deviation About Its Mean Value

One can write the total magnetization $\mathbf{M}$, as a mean magnetization $\mathbf{M_0}$ and a deviation from this mean value:

$$\mathbf{M} = \mathbf{M_0} + \Delta \mathbf{M} \; ; \; \mathbf{M_0} = const \tag{12}$$

For small deviations,

$$\frac{|\Delta \mathbf{M}|}{|\mathbf{M_0}|} = \frac{\Delta M}{M_0} \ll 1 \tag{13}$$

There is some interest in finding the EoM for the magnetization time derivative, $d\mathbf{M}/dt$. This requires one find the second derivative of (11), or

$$\frac{d^2 \mathbf{M}}{dt^2} \approx \gamma_e \frac{H_e a^2}{M} \frac{d}{dt} \{\mathbf{M} \times \nabla^2 \mathbf{M}\} \tag{14}$$

Performing the time derivation operation on the right-hand-side of (13), one finds

$$\frac{d^2 \mathbf{M}}{dt^2} \approx \left(\gamma_e \frac{H_e a^2}{M}\right)^2 \left\{\mathbf{M} \times (\mathbf{M} \times \nabla^4 \mathbf{M}) + \mathbf{M} \times (\mathbf{M} \times \nabla^2 \mathbf{M}) M \nabla^2 \left(\frac{1}{M}\right) + (\mathbf{M} \times \nabla^2 \mathbf{M}) \times \nabla^2 \mathbf{M}\right\} \tag{15}$$

Inserting (12) into this, generates the EoM for the magnetization devoation time derivative,

$$\frac{d^2 \Delta \mathbf{M}}{dt^2} = \left(\gamma_e \frac{H_e a^2}{M}\right)^2 \nabla^4 (\Delta \mathbf{M}) + (higher\ order\ terms\ in\ \Delta \mathbf{M}) \tag{16}$$

One can now understand the desire for looking at the second time derivative, namely to create something that might have harmonic or sinusoidal solutions, very familiar as propagating waves. Taking the solution as

$$\Delta \mathbf{M}_{ex}(t, \mathbf{r}) = \Delta \mathbf{M}_{ex0} e^{j(\omega t + \mathbf{k} \cdot \mathbf{r})} \tag{17}$$

For exchange magnetization, the radian frequency is obtained from the dispersion equation

$$\omega^2 = (\gamma_e H_e a^2)^2 k^4 \tag{18}$$

Whose solution is for the exchange deviation frequency

$$\omega_{ex,d} = \pm \gamma_e H_e a^2 k^2 \tag{19}$$

Because the gyromagnetic ratio $\gamma_e$ is negative, one takes the positive root solution for convenience,

$$\omega_{ex,d} = -\gamma_e H_e a^2 k^2 = \omega_{ex} a^2 k^2 \ ; \ \omega_{ex} = -\gamma_e H_e \tag{20}$$

Here $\omega_{ex}$ is the well recognized exchange frequency.

From (17) and (3), one expects the total magnetization deviation $\mathbf{M}_{Td}$ and total exchange field $\mathbf{H}_{Tex}$ can be expressed as Fourier series. By (12),

$$\mathbf{M}_T = \mathbf{M}_0 + \sum_\mathbf{k} \Delta \mathbf{M}_{ex} \tag{21}$$

so that

$$\mathbf{M}_{Tex,d} = \sum_\mathbf{k} \Delta \mathbf{M}_{ex} = \sum_\mathbf{k} \Delta \mathbf{M}_{ex}(t, \mathbf{r}) = \sum_\mathbf{k} \Delta \mathbf{M}_{ex0} e^{j(\omega t + \mathbf{k} \cdot \mathbf{r})} \tag{22}$$

The **k** vector summation can include the null point **k** = 0. However, it becomes apparent that this null momentum space vector does not contribute to $\mathbf{H}_{Tex}$:

$$\mathbf{H}_{Tex} = H_e \frac{a^2}{M}\nabla^2 \mathbf{M}_{Tex,d} = -H_e \frac{a^2}{M}\sum_\mathbf{k} k^2 \Delta \mathbf{M}_{ex0} e^{j(\omega t + \mathbf{k}\cdot\mathbf{r})} \tag{23}$$

### Complete Magnetodynamics Equation Using Total H Field

Of course, the complete EoM for **M**, for our magnetyodynamics equation in (8), will require the total H field, and that is given by

$$\mathbf{H}_T = \mathbf{H}_{internal} = \mathbf{H}_{ext} + \mathbf{H}_{ani} + \mathbf{H}_{ex} + \mathbf{H}_{d,\,static} + \mathbf{H}_{d,\,dynam} + \mathbf{H}_{rf} \tag{24}$$

Here $\mathbf{H}_i = \mathbf{H}_{internal}$, $\mathbf{H}_{ext}$, $\mathbf{H}_{ani}$, $\mathbf{H}_{d,\,static}$, $\mathbf{H}_{d,\,dynam}$, $\mathbf{H}_{rf}$, are, respectively, the internal, external or applied, anisotropy, depolarization or demagnetization static, depolarization or demagnetization dynamic, and RF magnetic fields. $\mathbf{H}_{ex}$ has already been determined, and from it and $\mathbf{M}_{Tex,d}$ using Maxwell's equations (taking displacement D field time variation as small and employing the magnetic material constitutive relationship),

$$\mathbf{H}_{d,\,dynam} = -\sum_\mathbf{k} \frac{\mathbf{k}(\mathbf{k}\cdot\mathbf{M_k})}{k^2} e^{j(\omega t + \mathbf{k}\cdot\mathbf{r})} \tag{25}$$

Expression for $\mathbf{H}_{d,\,static}$ is the usual expression,

$$\mathbf{H}_{d,\,static} = -\bar{\bar{\mathrm{N}}}\cdot\mathbf{M} \tag{26}$$

where $\bar{\bar{\mathrm{N}}}$ is the usual static depolarization tensor.

The magnetodynamics equation for magnetization can now be written as

$$\frac{d\mathbf{M}}{dt} = \gamma_e \mathbf{M}\times\mathbf{H}_T = \gamma_e \mathbf{M}\times\mathbf{H}_i \tag{27}$$

Lump the external field applied and the anisotropy fields together for compactness,

$$\mathbf{H}_{e,a} = \mathbf{H}_{ext} + \mathbf{H}_{ani} \tag{28}$$

Next define the shifted or modified $\mathbf{H}_{e,a}$ field as

$$\mathbf{H}^m_{e,a\,\mathbf{k}} = \mathbf{H}_{e,a} + \mathbf{H}_{ex} + \mathbf{H}_{d,\,static} \tag{29}$$

so that is components are

$$H^{xm}_{e,a\,\mathbf{k}} = H^x_{e,a} + N_x M^x_0 + \sum_\mathbf{k} H_e a^2 k^2 \frac{M^x_\mathbf{k}}{M} \tag{30a}$$

$$H_{e,a\,\mathbf{k}}^{ym} = H_{e,a}^y + N_y M_0^y + \sum_{\mathbf{k}} H_e a^2 k^2 \frac{M_{\mathbf{k}}^y}{M} \quad (30\text{b})$$

$$H_{e,a\,\mathbf{k}}^{zm} = H_{e,a}^z + N_z M_0^z + \sum_{\mathbf{k}} H_e a^2 k^2 \frac{M_{\mathbf{k}}^z}{M} \quad (30\text{c})$$

in a rectilinear tensor system for static demagnetization factor $\bar{\bar{N}}$. The $\mathbf{H}_{d,\,dynam}$ term in (24) does not contribute, which is why it did not appear in the final (30) formulas. And one did not have to turn on the last $\mathbf{H}_{rf}$ term to determine the spin wave

The EoM in $1\omega$ is by (27),

$$j\omega \begin{bmatrix} M_x(\omega) \\ M_y(\omega) \\ M_z(\omega) \end{bmatrix} = \gamma_e \begin{pmatrix} [M_y(\omega) H_{e,a\,\mathbf{k}}^{zmT}(0) - M_z(\omega) H_{e,a\,\mathbf{k}}^{ymT}(0)] \\ [M_z(\omega) H_{e,a\,\mathbf{k}}^{xmT}(0) - M_x(\omega) H_{e,a\,\mathbf{k}}^{zmT}(0)] \\ [M_x(\omega) H_{e,a\,\mathbf{k}}^{ymT}(0) - M_y(\omega) H_{e,a\,\mathbf{k}}^{xmT}(0)] \end{pmatrix} \quad (31)$$

Here the compacted components $H_{e,a\,\mathbf{k}}^{imT}$, $= x, y, z$, with included angular variations, are

$$H_{e,a\,\mathbf{k}}^{xmT} = H_{e,a\,\mathbf{k}}^{xm} - \sum_{\mathbf{k}} \cos(\phi)\sin(\theta)\cos(\theta) M_{\mathbf{k}} \quad (32\text{a})$$

$$H_{e,a\,\mathbf{k}}^{ymT} = H_{e,a\,\mathbf{k}}^{ym} - \sum_{\mathbf{k}} \sin(\phi)\sin(\theta)\cos(\theta) M_{\mathbf{k}} \quad (32\text{b})$$

$$H_{e,a\,\mathbf{k}}^{zmT} = H_{e,a\,\mathbf{k}}^{zm} - [\sin^2(\theta) - 1] M_0 \quad (32\text{c})$$

Magnetization components, or the eigenvectors may be determined, and also the dispersion relationship, which is

$$\omega_{spin} = \pm \left[ H_{e,a}^z + (N_y - N_z) M_0 - \sum_{\mathbf{k}} H_e a^2 k^2 + (\sin^2(\theta) - 1) M_0 \right]^{1/2} \times$$
$$\left[ H_{e,a}^z + (N_x - N_z) M_0 - \sum_{\mathbf{k}} H_e a^2 k^2 + (\sin^2(\theta) - 1) M_0 \right]^{1/2} \quad (33\text{a})$$

For a single $k$ value, obtaining $\omega_k$, one has

$$\omega_k = \pm \left[ H_{e,a}^z + (N_y - N_z) M_0 - H_e a^2 k^2 + (\sin^2(\theta) - 1) M_0 \right]^{1/2} \times$$
$$\left[ H_{e,a}^z + (N_x - N_z) M_0 - H_e a^2 k^2 + (\sin^2(\theta) - 1) M_0 \right]^{1/2} \quad (33\text{b})$$

Compare this to the Kittel resonance frequency form,

$$\omega_{res}^{Kittel} = \pm \left[ H_{e,a}^z + (N_y - N_z) M_0 \right]^{1/2} \left[ H_{e,a}^z + (N_x - N_z) M_0 \right]^{1/2} \quad (34)$$

which relates to spin precession due to a RF small signal external driving force, versus an internal driving force, say due to thermal agitation yielding our $\omega_k$ . That is why the equations are identical if one sets $\theta = \pi/2$ if one turns off the exchange effect, or sets $k = 0$.

### Magnetodynamics Equation for Twice the Spin Wave Frequency

On the left-hand-side of the **M** magnetodynamics torque equation, one can list the frequencies contributing in a list,

$$0 \cdot \omega_\mathbf{k}, \ 1 \cdot \omega_\mathbf{k}, \ 2 \cdot \omega_\mathbf{k}, \ 3 \cdot \omega_\mathbf{k}, \ 4 \cdot \omega_\mathbf{k}, \ \cdots \tag{35}$$

Now it is well known, that beyond a certain power level, spin waves may be launched at half the frequency of the RF signal frequency. That is,

$$\omega_\mathbf{k} = \frac{\omega_{rf}}{2} \tag{36}$$

We have observed the onset of this spin wave – electromagnetic interaction. This interaction is viewed as a macroscopic phenomena. On a more granular, or quantum level, the interaction is between magnons and photons. The EoM or the MDE is written down as, in vector form as

$$j2\omega_\mathbf{k}\mathbf{M}(2\omega_\mathbf{k}) = \gamma_e \mathbf{M}(0 \cdot \omega_\mathbf{k}) \times \mathbf{H}_{e,a\ \mathbf{k}}^{mT}(\omega_{rf}) \tag{37}$$

Consider the magnetization approximately saturated and in the z-direction, such that

$$\mathbf{M} \approx M_z \hat{z} = M_0 \hat{z} \tag{38}$$

Placing (38) into (37) yields

$$j2\omega_\mathbf{k}\mathbf{M}(2\omega_\mathbf{k}) = \gamma_e M_0 \hat{z} \times \mathbf{H}_{e,a\ \mathbf{k}}^{mT}(\omega_{rf}) \tag{39}$$

Inserting the vector forms into (37) generates

$$j2\omega_\mathbf{k} \begin{bmatrix} M_\mathbf{k}^x(2\omega_\mathbf{k}) \\ M_\mathbf{k}^y(2\omega_\mathbf{k}) \\ M_\mathbf{k}^z(2\omega_\mathbf{k}) \end{bmatrix} = \gamma_e M_0 \hat{z} \times \begin{bmatrix} H_{e,a}^{xmT}(\omega_{rf}) \\ H_{e,a}^{ymT}(\omega_{rf}) \\ H_{e,a}^{zmT}(\omega_{rf}) \end{bmatrix} \tag{40}$$

Performing the cross-product multiplication on the right-hand-side of (40),

$$j2\omega_\mathbf{k} \begin{bmatrix} M_\mathbf{k}^x(2\omega_\mathbf{k}) \\ M_\mathbf{k}^y(2\omega_\mathbf{k}) \\ M_\mathbf{k}^z(2\omega_\mathbf{k}) \end{bmatrix} = \gamma_e M_0 [H_{e,a}^{xmT}(\omega_{rf})\hat{y} - H_{e,a}^{ymT}(\omega_{rf})\hat{x}] \tag{41}$$

Writing (41) out, by vector element row, and equating to the appropriate right-hand-sided term, one obtains

$$j2\omega_{\mathbf{k}} M_{\mathbf{k}}^x(2\omega_{\mathbf{k}}) = -\gamma_e M_0 H_{e,a}^{ymT}(\omega_{rf}) \tag{42a}$$

$$j2\omega_{\mathbf{k}} M_{\mathbf{k}}^y(2\omega_{\mathbf{k}}) = +\gamma_e M_0 H_{e,a}^{xmT}(\omega_{rf}) \tag{42b}$$

Here $H_{e,a}^{xmT}(\omega_{rf})$ and $H_{e,a}^{ymT}(\omega_{rf})$ are the time varying components of the total **H** magnetic field, with frequency variation $\omega_{rf}$. If one assumes for the time being that it is the dominant part of each of these field components, then one has

$$H_{e,a}^{xmT}(\omega_{rf}) = h_{rf}^x e^{j\omega_{rf}t} \quad ; \quad H_{e,a}^{ymT}(\omega_{rf}) = h_{rf}^y e^{j\omega_{rf}t} \tag{43}$$

for spatially uniform applied RF magnetic field from our electromagnetic input signal to the device under test. The magnetization components are

$$M_{\mathbf{k}}^x(2\omega_{\mathbf{k}}) = M_{\mathbf{k}0}^x e^{j(2\omega_{\mathbf{k}}t+\mathbf{k}\cdot\mathbf{r})} \quad ; \quad M_{\mathbf{k}}^y(2\omega_{\mathbf{k}}) = M_{\mathbf{k}0}^y e^{j(2\omega_{\mathbf{k}}t+\mathbf{k}\cdot\mathbf{r})} \tag{44}$$

When (43) and (44) are put into (42), the only reasonable way to cancel out the spatial variation in the magnetic material sample with the uniform applied magnetic field, is to set $\mathbf{r} = \mathbf{0}$, or for the magnitude, $r = 0$. One has

$$j2\omega_{\mathbf{k}} M_{\mathbf{k}0}^x e^{j(2\omega_{\mathbf{k}}t+\mathbf{k}\cdot\mathbf{0})} = -\gamma_e M_0 h_{rf}^y e^{j\omega_{rf}t} \tag{45a}$$

$$j2\omega_{\mathbf{k}} M_{\mathbf{k}0}^y e^{j(2\omega_{\mathbf{k}}t+\mathbf{k}\cdot\mathbf{0})} = +\gamma_e M_0 h_{rf}^x e^{j\omega_{rf}t} \tag{45a}$$

By (45), invoking (36), to cancel out the remaining exponential time factors,

$$j2\omega_{\mathbf{k}} M_{\mathbf{k}0}^x = -\gamma_e M_0 h_{rf}^y \quad ; \quad j2\omega_{\mathbf{k}} M_{\mathbf{k}0}^y = +\gamma_e M_0 h_{rf}^x \tag{46}$$

From (46), one may solve for the applied external field strength, whose square is proportional to the power required to activate the critical field strength needed to launch spin waves and limit the signal strength. Such limiting is necessary in systems so that the components are not damaged or destroyed, or the desired processing not overwhelmed by unwanted sources.

$$h_{rf}^x = j \frac{\omega_{rf}}{\gamma_e} \frac{M_{\mathbf{k}0}^y}{M_0} \quad ; \quad h_{rf}^y = -j \frac{\omega_{rf}}{\gamma_e} \frac{M_{\mathbf{k}0}^x}{M_0} \tag{47}$$

Assigning the gyrating resonance frequency as

$$\omega_0 = \gamma_e M_0 \tag{48}$$

(47) becomes

$$h_{rf}^x = j \frac{\omega_{rf}}{\omega_0} M_{\mathbf{k0}}^y \quad ; \quad h_{rf}^y = -j \frac{\omega_{rf}}{\omega_0} M_{\mathbf{k0}}^x \tag{49}$$

Thus their magnitudes are

$$\left|h_{rf}^x\right| = \frac{\omega_{rf}}{\omega_0} M_{\mathbf{k0}}^y \quad ; \quad \left|h_{rf}^y\right| = \frac{\omega_{rf}}{\omega_0} M_{\mathbf{k0}}^x \tag{50}$$

Finally taking the major component of magnetization as that in the z-direction, $M_{\mathbf{k}}^z$, and viewing the transverse components utilized in (50) as much smaller, that is,

$$M_{\mathbf{k0}}^x, M_{\mathbf{k0}}^y \approx \frac{1}{4\pi} \Delta H_{\mathbf{k}} \tag{51}$$

Equations (50) covert into a single statement,

$$\left|h_{rf}\right| \sim \frac{\omega_{rf}}{\omega_0} \Delta H_{\mathbf{k}} \tag{52}$$

And noting that this RF field ought to be the critical field at which spin waves are launched in substantial amount,

$$h_{critical} \approx \left|h_{rf}\right| \tag{53}$$

Which means that

$$h_{critical} \sim \frac{\omega_{rf}}{\omega_0} \Delta H_{\mathbf{k}} \tag{54}$$

Equation (53) says that the power level required to launch the spin waves and create the limiting action, can be maintained constant if

$$\frac{\omega_{rf}}{\omega_0} \Delta H_{\mathbf{k}} = \text{const} \tag{55}$$

Recognizing that the linewidths for different k values may be similar in size, one can drop that subscript in to find

$$\frac{\omega_{rf}}{\omega_0} \Delta H = \text{const} \tag{56}$$

For magnetic materials with similar sized linewidths, this equation reduces to its most simple form:

$$\frac{\omega_{rf}}{\omega_0} = \text{const} \tag{57}$$

Thus, as one raises the frequency of operation of the limiter, the material's magnetization must concomitantly tract this increase. Therefore, to go from the 5 – 15 GHz regime to the 40 - 60 GHz regime, supposing $M$ values were on the order of 1000 – 2000 Oe in the lower regime, $M$ must rise by a factor of about (40 – 60)/(5 – 15) = 4. Therefore, $M$ must be in the range of 4000 – 8000 Oe. So ones looks to newer much higher magnetization materials with excellent and small linewidths.

## Conclusions

In this letter it has been shown how to relate the critical magnetic field determining the onset of nonlinear spin wave – electromagnetic interaction, to key material parameters of the materials. It may be entirely possible to develop new magnetic materials in the future for considerably higher frequency operation. High quality magnetic materials with excellent low loss behavior, characterized by small linewidths, will still be required, and now with much enhanced static magnetization. Although, conventionally this can be achieved using large electromagnets to apply the field, or powerful fixed magnets, obviously such solutions obtained in an experimental laboratory test setup or puck type magnetic RF device, the way perhaps in the future is with self-biased magnetic materials using newly developed deposition and processing technologies [38] – [42].

What is needed for RF electromagnetic (EM) devices is the ability to manipulate the EM waves, and this adds on a further requirement that the materials possess the qualities of low loss seen in dielectric materials commonly used as fillers or substrates for microwave and millimeter wave and higher frequency RF circuits. Thus, although Takao Suzuki [43] in his seminar in 2015 mentions many interesting magnetic materials, these seem mostly suited for magnets, and so tolerating conductive metallic materials is acceptable. Not so with the usual RF component needs. However, the attention directed in his seminar towards high anisotropy fields may be very appropriate for obtaining the self-biasing high magnetizations desired for not only circulator materials but also materials for limiters at the much higher frequencies entertained in our treatment. Some perspective might be gained for the reader by referring to a recent article on microwave integrated magnetics [44]. One also notes current interest in the spin wave – electromagnetic wave interaction in two other recent contributions, [45], [46] and [47]. The first addresses, through a phenomenological model of non-linear loss in ferrimagnetic materials, frequency-selective RF limiter behavior [45]. [46]. The second [47], develops a spin wave – electromagnetic physics based coupling nonlinear circuit model for frequency-selective RF limiters.

## **Acknowledgements**

This work was supported by the Office of Naval Research. Would like to thank our group working on the fabrication and measurements of limiter devices using magnetic materials. This includes Scooter Johnson (MSTD) on material growth as well as measurement, Sanghoon Shin (TEWD) in RF circuit fabrication using microstrip/stripline technologies, Sayed Quadri (MSTD) in materials growth area, and Hans Haucke (TEWD) in experimental measurement assistance.

## **References**

[1] J. Brown, "Ferromagnetic limiters," Micro. J., vol. 4, pp. 74–79, Nov. 1961.
[2] D. R. Jackson and R. W. Orth, "A frequency-selective limiter using nuclear magnetic resonance," Proc. IEEE, vol. 55, no. 1, pp.36–45, Jan. 1967.
[3] R. W. Orth, "Frequency-selective limiters and their application," IEEE Trans. Electromagn. Compat., vol. EMC-10, no. 2, pp.273–283, Jun. 1968.
[4] P. R. Emtage and S. N. Stitzer, "Interaction of signals in ferromagnetic microwave limiters," IEEE Trans. Micro. Th. Tech., vol. MTT-25, no. 3, pp. 210–213, Mar. 1977.
[5] A. J. Giarola, "A review of the theory characteristics and operation of frequency selective limiters," Proc. IEEE, vol. 67, no. 10, pp. 1380–1396, Oct. 1979.
[6] S. Stitzer, "Frequency selective microwave power limiting in thin YIG films," IEEE Trans. Magn., vol. MAG-19, no. 5, pp. 1874–1876, Sep. 1983.
[7] J. D. Adam and S. N. Stitzer, "Frequency selective limiters for high dynamic range microwave receivers," IEEE Trans. Micro. Th.Tech., vol. 41, no. 12, pp. 2227–2231, Dec. 1993.
[8] S. N. Stitzer, "Spike leakage and suppression in frequency selective limiters," in IEEE MTT-S Int. Micro. Symp. Dig., vol. 2, Jun. 2000, pp. 901–904.
[9] J. D. Adam, L. E. Davis, G. F. Dionne, E. F. Schloemann, and S. N. Stitzer, "Ferrite devices and materials," IEEE Trans. Micro. Th. Tech., vol. 50, no. 3, pp. 721–737, Mar. 2002.
[10] J. D. Adam, "Mitigate the interference: Nonlinear frequency selective ferrite devices," IEEE Micro. Mag., vol. 15, no. 6, pp. 45–56, Sep./Oct. 2014.
[11] S. M. Gillette, M. Geiler, J. D. Adam, and A. L. Geiler, "Ferrite-based reflective-type frequency selective limiters," in Proc. IEEE MTT-S Int. Micro. Workshop Ser. Adv. Mater. Processes RF THz Appl. (IMWSAMP), Ann, Arbor, MI, USA, Jul. 2018, pp. 1–3.
[12] H. Suhl, "The nonlinear behavior of ferrites at high microwave signal levels," Proc. IRE, vol. 44, no. 10, pp. 1270–1284, Oct. 1956.
[13] F. R. Morgenthaler, "Survey of ferromagnetic resonance in small ferromagnetic ellipsoids," J. Appl. Phys., vol. 31, no. 5, p. S95, 1960.
[14] E. Schlömann, J. J. Green, and U. Milano, "Recent developments in ferromagnetic resonance at high power levels," J. Appl. Phys., vol. 31, no. 5, p. S386, 1960.
[15] M. Cheng and C. E. Patton, "Spin wave instability processes in ferrites," in Nonlinear Phenomena and Chaos in Magnetic Materials, P. E.Wigen, Ed. Singapore: World Scientific, 1992.
[16] D. D. Stancil, A. Prabhakar, Spin Waves: Theory and Applications. New York, NY, USA: Springer, 2009. D. D. Stancil, Theory of Magnetostatic Waves. New York-USA & Berlin-Germany, Springer-Verlag, 1993.
[17] V. Laur , R. Lebourgeois, E. Laroche, J.L. Mattei, P. Quéffélec, J.P. Ganne and G. Martin, "Low-Loss Millimeter-Wave Self-Biased Circulators: Materials, Design and Characterization," SPDC Sopace Proceed. 2016. (All authors from France) (1)Lab-STICC, University of Brest, Brest, France
[18] V. Laur, R. Lebourgeois, E. Laroche, J.L. Mattei, P. Quéffélec, J.P. Ganne, G. Martin, "Study of a low-loss self-biased circulator at 40 GHz: influence of temperature," in Proc. IEEE Int. Micr. Symp., 2016.
[19] V. Laur, G. Vérissimo, P. Queffelec, L.A. Farhat, H. Alaaeddine, E. Laroche, G. Martin, R. Lebourgeois, J.P. Ganne, "Selfbiased Y-junction Circulators using Lanthanum- and Cobalt-substituted Strontium hexaferrites" IEEE Trans. Microwave Theory & Techn., vol. 63, no. 12, pp. 4376-4381, December 2015.

[20] V. Laur, G. Vérissimo, P. Quéffélec, L.A. Farhat, H. Alaaeddine, J.C. Reihs, E. Laroche, G. Martin, R. Lebourgeois and J.P. Ganne, "Modeling and characterization of self-biased circulators in the mm-wave range," in Proc. IEEE Int. Micr. Symp., 2015.
[21] Investigation of Ferromagnetic Resonance Shift in Screen-Printed Barium Ferrite/Samarium Cobalt Composites, IEEE Trans. Micro. Th. Tech. 76, 3230 – 3236, Aug. 2019.
[22] Takao Suzuki, High Magnetic Anisotropy Materials, Seminar, Center for Materials for Information Technology (MINT), Univ, Alabama, Feb. 6, 2015.
[23] J. Wang, A. Yang, Y. Chen, Z. Chen, A. Geiler, S.M. Gillette, V.G. Harris and C. Vittoria, "Self-biased Yjunction circulator at Ku band," IEEE Micro. Wireless Compon. Lett., vol. 21, no. 6, pp. 292-294, June 2011.
[24] B.K. O'Neil and J.L. Young, "Experimental investigation of a self-biased microstrip circulator," IEEE Trans. Micro. Th. & Tech., vol. 57, no. 7, pp. 1669-1674, July 2009.
[25] X. Zuo, H. How, S. Somu and C. Vittoria, "Self-biased circulator/isolator at millimeter wavelengths using magnetically oriented polycrystalline strontium M-type hexaferrite," IEEE Trans. Magnetics, vol. 39, no. 5, pp. 3160-3162, September 2003.
[26] S.A. Oliver, P. Shi, W. Hu, H. How, S.W. McKnight, N.E. McGruer, P.M. Zavracky and C. Vittoria, "Integrated self-biased hexaferrite microstrip circulators for millimeter-wavelength applications," IEEE Trans. Microwave Th. & Tech., vol. 49, no. 2, pp. 385-387, February 2001.
[27] M.A. Tsankov and L.G. Milenova, "Design of self-biased hexaferrite waveguide circulators," Journal of Applied Physics, vol. 73, no. 10, pp. 7018-7020, May 1993.
[28] N. Zeina, H. How, C. Vittoria and R. West, "Self-biasing circul ators operating at Ka-band utilizing M-type hexagonal ferrites,"IEEE Trans. Magnetics, vol. 28, no. 5, pp. 3219-3221, September 1992.
[29] Han Cui, Zhi Yao, Yuanxun Ethan Wang, Coupling Electromagnetic Waves to Spin Waves: A Physics-Based Nonlinear Circuit Model for Frequency-Selective Limiters, IEEE Trans. Micro. Th. Tech. 67, 3221 – 3229, Aug. 2019.
[30] Anatoliy O. Boryssenko, Scott M. Gillette, Marina Y. Koledintseva, "Nonlinear Loss Model in Absorptive-Type Ferrite Frequency-Selective Limiters," IEEE Trans. Micro. Th. Tech. 67, 4871 – 4880, Aug. 2019.
[31] Ronald A. Soohoo, Theory and Applications of Ferrites, Englewood Cliffs, New Jersey, Prentice-Hall, 1960.
[32] C. M. Krowne, "Electromagnetic Distributions Demonstrating Asymmetry Using a Spectral Domain Dyadic Green's Function for Ferrite Microstrip Guided Wave Structures," IEEE Trans. Microwave Th. Tech., Vol. 53, pp. 1345 - 1361, Apr. 2005.
[33] C. M. Krowne, "Numerical Modelling of Circulators," under Ferrite Devices heading, in Encyclopedia of Electrical and Electronics Engineering, Ed, John G. Webster, Vol. 15, pp. 1 - 19, Wiley, 1999.
[34] C. M. Krowne, "CAD Using Dyadic Green's Functions and Finite Elements and Comparison to Experiment for Inhomogeneous Ferrite Microstrip Circulators," in Advances in Imaging and Electron Physics, Ed. Peter W. Hawkes, Vol. 106, pp. 97 - 184, Academic Press, 1999.
[35] C. M. Krowne, "Dyadic Green's Function Microstrip Circulator Theory for Inhomogeneous Ferrite with and without Penetrable Walls ," in Advances in Imaging and Electron Physics, Ed. Peter W. Hawkes, Vol. 103, pp. 151 - 275, Academic Press, 1998.

[36] C. M. Krowne, "Theory of the Recursive Dyadic Green's Fu nction for Inhomogeneous Ferrite Canonically Shaped MicrostripCirculators," in Advances in Imaging and Electron Physics, Ed. Peter W. Hawkes, Vol. 98, pp. 77 - 321, Academic Press, 1996.

[37] C. M. Krowne and R. E. Neidert, "Theory and Numerical Calculations For Radially Inhomogeneous Circular Ferrite Circulators,"IEEE Trans. Microwave Theory & Tech., Vol. 44, no. 3, pp. 419-431, March 1996.

[37] S. D. Johnson, E. A. Patterson, J. Xing, et al., "Design and Implementation of a Magnetic Press System for Creating Magnetically Oriented Barium Hexaferrite Pucks," IEEE Trans. Magn. 55, 2801406, 2019.

[38] S. D. Johnson, D.-S. Park, S. B. Quadri, E. P. Gorzkowski, "Formation of magnetically-oriented barium hexaferrite films by aerosol deposition," J. Magn. Magn. Mater. 479, 156 – 160, 2019.

[39] Scooter D. Johnson, Christopher M. Gonzalez, Virginia Anderson, Zachary Robinson, Harvey S. Newman, Sanghoon Shin and Syed B. Qadri, "Magnetic and structural properties of sintered bulk pucks and aerosol deposited films of Ti-doped barium hexaferrite for microwave absorption applications," J. Appl. Phys. **122**, 024901 (2017).

[40] Scooter D. Johnson, Evan R. Glaser, Shu-Fan Cheng, Jennifer Hite, "Dense nanocrystalline yttrium iron garnet films formed at room temperature by aerosol deposition," Mater Res. Bull. 76, 365- 369, 2016.

[41] Johnson, S. D., Glaser, E. R., Kub, F. J., Eddy, Jr., C. R. Formation of Thick Dense Yttrium Iron Garnet Films Using Aerosol Deposition. *J. Vis. Exp.* (99), e52843, doi:10.3791/52843 (2015).

[42] Scooter D. Johnson, Evan R. Glaser, Shu-Fan Cheng, Fritz J. Kub, and Charles R. Eddy, Jr., "Characterization of as-deposited and sintered yttrium iron garnet thick films formed by aerosol deposition," Appl. Phys. Exp. 7, 035501 (2014).

[43] Takao Suzuki, "High Magnetic Field Anisotropy Materials," MINT Center Seminar, University of Alabama, 2015.

[44] William Palmer, David Kirkwood, Steve Gross , Michael Steer , Harvey S. Newman, Scooter Johnson, "A Bright Future for Integrated Magnetics," IEEE Microwave Magazine, 36 – 50, June 2019.

[45] Anatoliy O. Boryssenko, Scott M. Gillette, and Marina Y. Koledintseva , Nonlinear Loss Model in Absorptive-Type Ferrite Frequency-Selective Limiters," IEEE Trans. Microwave Theory & Tech., Vol. 67, no. 12, pp. 4871-4880, Dec. 2019.

[46] A. Boryssenko, S. Gillette, and M. Koledintseva, "A phenomenological model of non-linear loss in ferrimagnetic frequency-selective limiters," IEEE MTT-S Int. Microw. Symp. Dig., , pp. 595–598, Boston, MA, USA, Jun. 2019.

[47] Han Cui, Zhi Yao, Yuanxun Ethan Wang, Coupling Electromagnetic Waves to Spin Waves: A Physics-Based Nonlinear Circuit Model for Frequency-Selective Limiters," IEEE Trans. Micro. Th. Tech. 67, no. 8, pp 3221 – 3229, Aug. 2019.